\documentclass{svproc}
\usepackage[utf8]{inputenc}

\usepackage{url}

\usepackage{natbib}
\bibpunct{(}{)}{;}{a}{}{,}%
\usepackage{graphicx}
\usepackage{amsmath,amssymb}
\usepackage{multirow}
\usepackage{longtable}

\usepackage{mathtext}
\usepackage{journals}
\usepackage{latexsym}
\usepackage{url}
\usepackage{listings}
\usepackage{graphicx}
\usepackage{euscript}
\usepackage{amsmath}
\usepackage{subcaption}
\usepackage[utf8]{inputenc}
\usepackage{multirow}
\usepackage[usenames]{color}
\usepackage{amsmath}
\usepackage{hyperref}
\usepackage{caption,setspace}

\let\oldAA\AA
\renewcommand{\AA}{\text{\normalfont\oldAA}}

\usepackage[b5paper]{geometry}
\begin{document}
	\mainmatter              
	\title{Optical spectroscopic observations of intermediate-mass black holes and their host galaxies: the $M_{BH}-\sigma_*$ relation.}
	\titlerunning{$M_{BH}-\sigma_*$ relation for intermediate-mass black holes.}  
	%
	\author{Vladimir Goradzhanov\inst{1,3} \and Igor Chilingarian\inst{1,2}
		Ivan Katkov\inst{1,4,5} \and Kirill Grishin\inst{1}
		\and Victoria Toptun\inst{1,3} \and Ivan Kuzmin\inst{1,3}
		\and Mariia Demianenko\inst{1,6,7}}
	\authorrunning{Goradzhanov et al.} 
	%
	\tocauthor{Vladimir Goradzhanov, Igor Chilingaryan, Ivan Katkov,
	           Kirill Grishin, Victoria Toptun, Ivan Kuzmin, Mariia Demianenko}
	\institute{Sternberg Astronomical Institute, M.V.Lomonosov Moscow State University,\\
	    \and
	    Center for Astrophysics -- Harvard and Smithsonian (USA),\\
		\and
		Department of Physics, M.V. Lomonosov Moscow State University,\\
		\and
		New York University Abu Dhabi (UAE),\\
		\and
		Center for Astro, Particle, and Planetary Physics, NYU AD (UAE)\\
		\and
		Moscow Institute of Physics and Technology (National Research University), \\
		\and
		HSE University}
	\maketitle 
	
\begin{abstract}
Intermediate-mass black holes (IMBHs; $M_{BH} <2*10^{5} M_{\odot}$) in galaxy centers are cruciel for painting a coherent picture of the formation and growth of supermassive black holes (SMBHs). 
Using \textit{Big Data} analysis, we identified 305 IMBH candidates for IMBH and 1623 candidates of `light-weight' SMBHs ($2 * 10^{5} M_{odot} < M_{BH} <10^{6} M_{\odot}$). For 35 host galaxies from this combined sample with the X-ray-confirmed active galactic nuclei (AGN) we collected and analyzed optical spectroscopic observations. These data show that bulge stellar velocity dispersions ($\sigma_*$) lie in the range of 24$\dots$118~km/s and do not follow the correlation with $M_{BH}$ established by larger SMBHs indicating that in the $10^{5}-10^{6} M_{\odot}$ range the accretion is the prevailing BH growth channel.
\end{abstract}

\section{Introduction}

Currently observed black holes are categorized by mass into three types: stellar mass black holes ($4<M_{BH}<100 M_{\odot}$), supermassive black holes (SMBH; $M_{BH}>10^5 M_{\odot}$), and intermediate mass black holes (IMBH; $100<M_{BH}<10^5 M_{\odot}$). The growth of central SMBHs and the way they affect their host galaxies remains a highly controversial issue in modern astrophysics. From theoretical perspectives, a black hole seed with a starting mass of $50 M_{\odot}$ cannot grow to a billion solar mass SMBH via accretion in $\sim$1~Gyr (or $z>7$ corresponding to record holding high-redshift quasars), because the accretion rate is limited by the Eddington luminosity ($L_{Edd}$). If this is indeed the main SMBH growth channel, then `undergrown' seeds will form a population of IMBHs.

Here we adopt the limit $ M_{\mathrm {BH}} <2\cdot 10^{5} M_{\odot}$ as the threshold mass for what we call an IMBH, because it corresponds to +1$\sigma$ systematic uncertainty in virial mass estimate \citep{reines13, Chilingarian+18} from broad $H\alpha$.

Central massive black holes in both galaxies with active and inactive nuclei follow rather tight correlations with the stellar velocity dispersion $\sigma_{*}$ of the bulge and the total stellar mass of the bulge $M_{\rm bulge}$ \citep{FM00, vandenBosch16}, which is interpreted as a co-evolution of the bulge and SMBH via galaxy mergers. At the same time, SMBHs can gain mass via accretion \citep{2012Sci...337..544V} at any moment during the cosmic time. It is not yet fully known whether these relations hold for IMBH because of observational difficulties and small samples of known IMBHs.

In 2018, the largest to-date sample of 305 IMBH candidates was identified by \citet{Chilingarian+18} using data mining in the Reference Catalog of Spectral Energy Distribution \citep{RCSED} database of almost 1 million spectra of galaxies from the SDSS Survey \citep{SDSS_DR7}.  Some of them were later observed with intermediate-resolution optical spectrographs MagE (6.5-m Magellan telescope) and RSS (10-m Southern African Large Telescope) to refine stellar velocity dispersions and improve $M_{BH}$ measurements. Here we report preliminary results of this project and also present $\sigma_{*}$ measurements for a sample of `light-weight' SMBHs ($<10^6 M_{\odot}$) populating the bottom part of the $M_{BH}-\sigma_{*}$ relation.

\section{Spectral data analysis}

\begin{figure}
\begin{tabular}{rl}
\includegraphics[trim=4cm 5.5cm 4cm 5cm, clip, height=1.0cm, width=8.0cm]{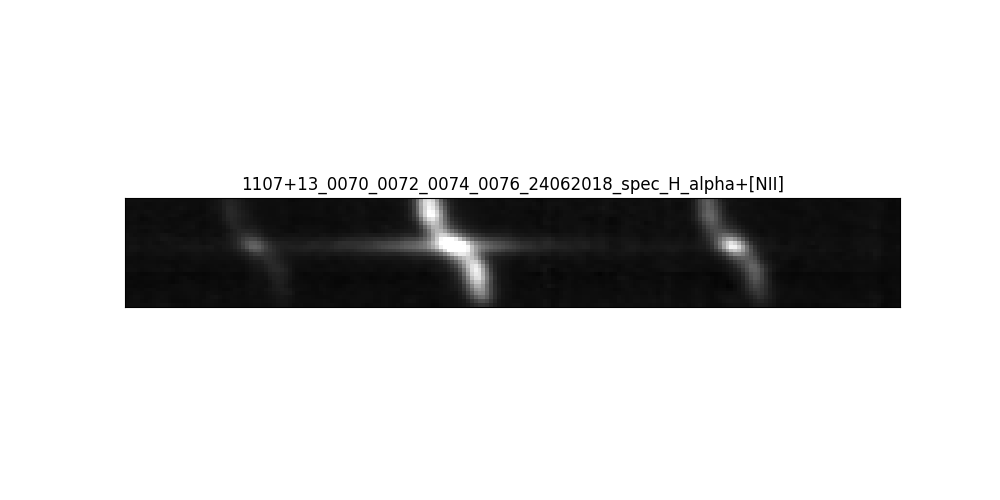} & \multirow[c]{4}{*}{\includegraphics[trim=8cm -1.5cm 0.5cm 1.0cm, clip, width=6cm,height=4.5cm]{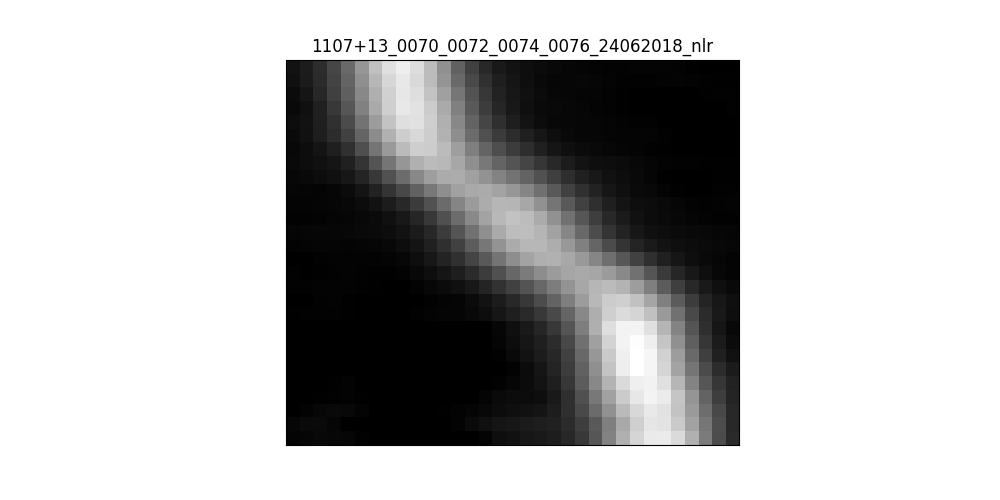}} \\
\includegraphics[trim=4cm 5.5cm 4cm 5cm, clip, height=1.0cm, width=8.0cm]{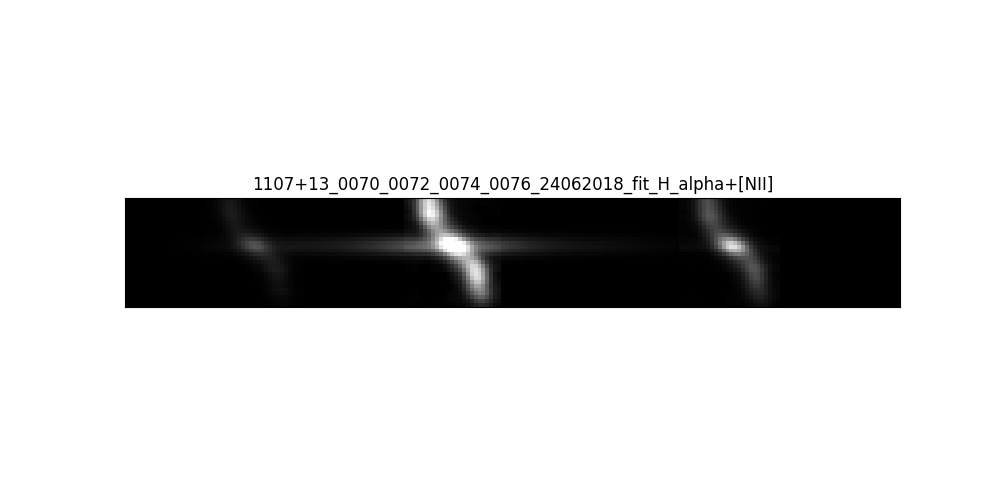} & \\
\includegraphics[trim=4cm 5.5cm 4cm 5cm, clip, height=1.0cm, width=8.0cm]{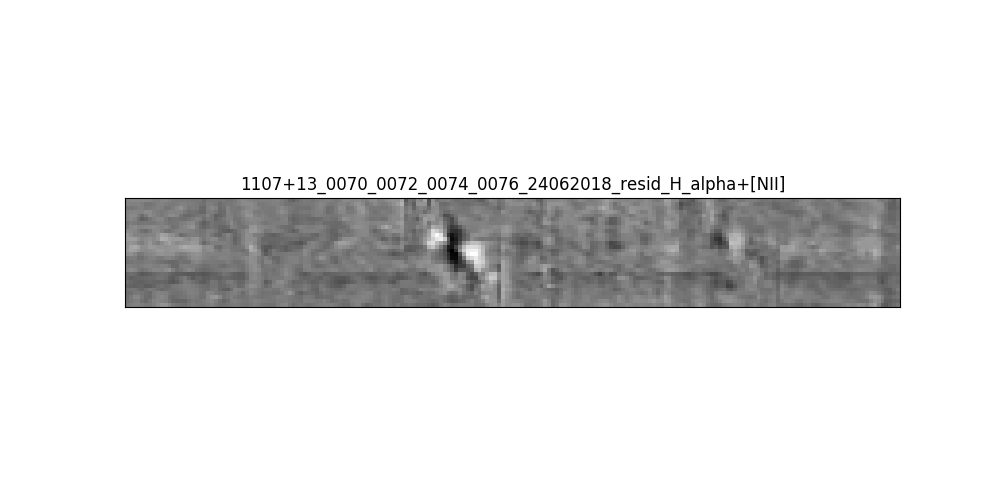} & \\
\includegraphics[trim=4cm 5.5cm 4cm 5cm, clip, height=1.0cm, width=8.0cm]{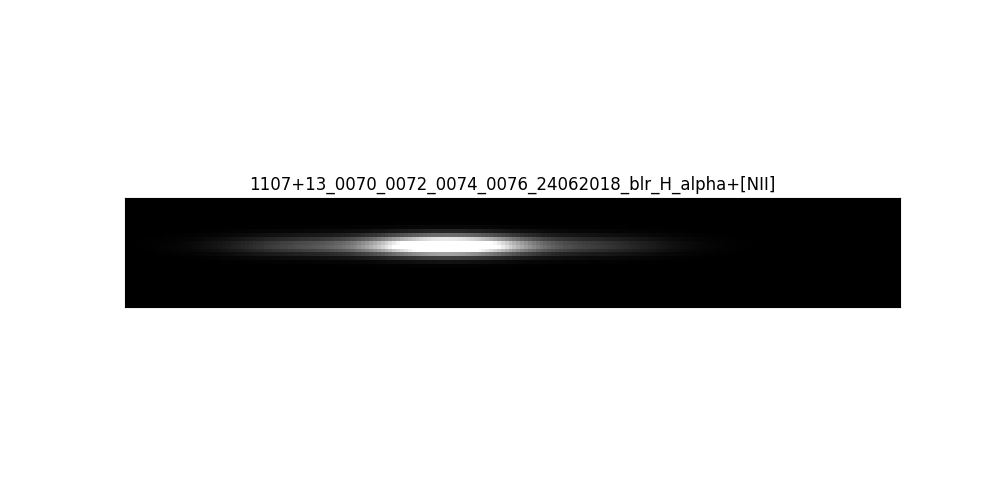} & \\
\end{tabular}
{\setstretch{1.0} 
\caption{Left panels top to bottom: the original 2d emission spectrum of the galaxy J110731.22+134712.9 in the $H\alpha$ region; the result of the fitting of emission lines; fitting residuals; the reconstructed broad line profile. Right panel: a 2D narrow-line profile of the $H\alpha$.\label{fig:2d_fit.png}}}
\end{figure}

\begin{figure}[htp!]
\begin{center}
\vskip -4mm
\includegraphics[width=0.53\linewidth]{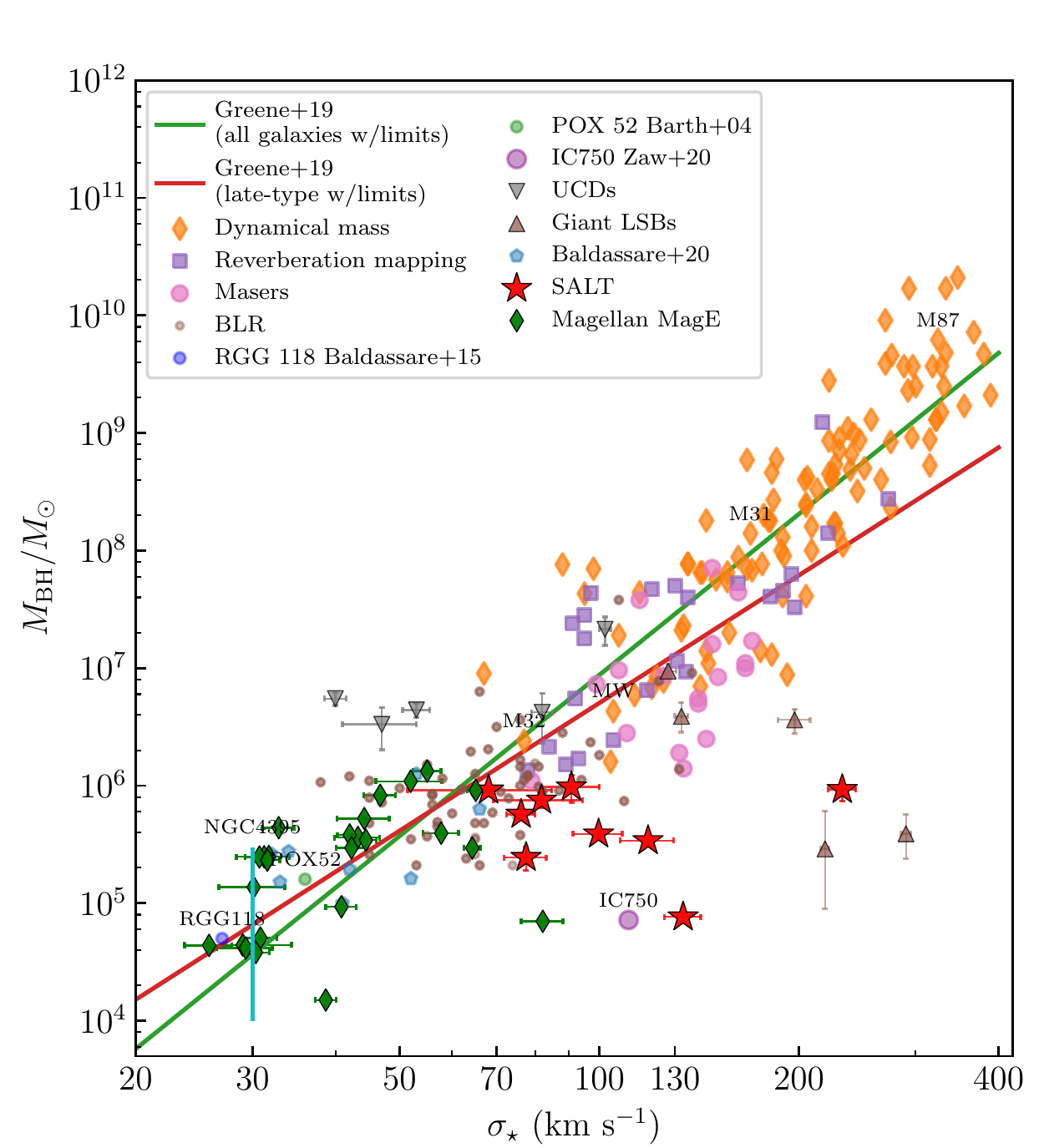}
{\setstretch{1.0} 
\caption{The $M_{BH}-\sigma_*$ relation featuring IMBHs and light-weight SMBHs. Only one IMBH has a mass estimate based on masers \citep{Zaw2020}; the rest of the data points are virial estimates from $H\alpha$. New data from MagE and SALT are shown by green diamonds and red stars respectively.}

\label{fig:M_sigma.pdf}}
\end{center}
\end{figure}

We obtained 30 spectra for 26 `light-weight' SMBHs using MagE and additional 9 spectra using RSS and reduced them with standard data processing pipelines. We fitted the reduced spectra with the {\sc NBursts} technique \citep{CPSA07,CPSK07} to determine the stellar continuum and measure stellar velocity dispersions as low as 20~km/s \citep{2020PASP..132f4503C}. Then we subtracted the best-fitting stellar population templates from the spectra and applied a novel method of a two-dimensional nonparametric reconstruction of the $H\alpha$ profile of emission spectra (see Fig.~\ref{fig:2d_fit.png}). The method yields a non-parametric narrow-line profile featuring narrow lines from the AGN and a position--velocity diagram for narrow lines originating from a star-forming disk of its host galaxy. It also returns the parameters of the broad component of $H\alpha$ and $H\beta$ formed in the broad-line region, as well as corresponding fluxes. The broad $H\alpha$ flux and width are then used for the virial $M_{BH}$ estimate using the calibration from \citet{reines13}: $M_{BH} = 3.72\cdot10^6 (\mathrm{FWHM}_{H\alpha}/10^3 \mathrm{km/s})^{2.06}\cdot(L_{H\alpha}/10^{42}\mathrm{erg/s})^{0.47}$. We rejected 7 galaxies from the MagE sample where broad components were not confidently detected. We also took into account slit losses assuming that the broad H$\alpha$ component comes from a point source and knowing the slit width.


\section{The $M_{BH}-\sigma$ scaling relation}

The $ M_{BH}-\sigma_{buldge} $ relation constructed from the measured values of $\sigma_*$ and $M_{BH}$ derived from MagE and SALT data is shown in Fig.~\ref{fig:M_sigma.pdf}. We can make a preliminary conclusion that light-weight SMBHs and IMBHs do not follow the relation established by more massive SMBHs and, hence, do not co-evolve with bulges of their host galaxies. Such situation can arise if the black hole mass growth in the low-mass regime is dominated by accretion rather than mergers. This result might have serious implications on the detectability of gravitational wave signals from merging IMBHs by the fortcoming LISA mission.
\vskip 2mm
\textbf{Acknowledgements.} This project is supported by the RScF Grant 17-72-20119.

\begingroup
\let\clearpage\relax
\bibliographystyle{aasjournal}
\bibliography{main}
\endgroup

\end{document}